# Circular motion and chaos bound of a charged particle near charged 4D Einstein-Gauss-Bonnet-AdS black holes

Jiayu Xie, Jie Wang, Bing Tang*

*Department of Physics, Jishou University, Jishou 416000, China*

**ABSTRACT**

The circular motion and chaos bound of one charged particle near 4D charged AdS black holes in the Einstein-Gauss-Bonnet gravity theory are analytically investigated. With the help of the Jacobian matrix, we construct the actual formula of the Lyapunov exponent for the charged particle, which satisfies the upper bound when it is localized at the event horizon. Further considering the Lyapunov exponent in the vicinity of the horizon and studying the 4D charged Einstein-Gauss-Bonnet-AdS black hole with different Gauss-Bonnet coupling coefficients, it is found that it has some specific values to determine whether a violation of chaos bound. The Lyapunov exponent for the circular motion of charged test particle is larger than the static equilibrium because of the appearance of angular momentum. We find that, with the increase of the Gauss-Bonnet coupling coefficient, the black hole gets closer to the extremal state and the bound is more easily violated. For various Gauss-Bonnet coupling coefficients, we obtain a corresponding range of the particle charge, in which the chaos bound is violated. Our results indicate that, as the Gauss-Bonnet coupling coefficient increases, the value of the particle charge violating the chaos bound is even smaller.

## 1. Introduction

Chaos is one of the most significant physical phenomena in those nonlinear dynamical systems. A system is supposed to be chaotic when trajectories of dynamics are highly sensitive to the initial conditions [1-4]. Sensitivity to the initial condition in the context of the boundary theory has been identified by the AdS/CFT correspondence [5-8], where the weak coupling case of the AdS aspect corresponds to the strong coupling limit of the CFT aspect. Hence, the non-perturbative quantity in CFT can be computed by employing the perturbative quantity in AdS due to the AdS/CFT correspondence. More significantly, the chaotic phenomenon in the quantum systems has been investigated by means of the AdS/CFT correspondence [9, 10]. In the quantum nonlinear dynamic system, an out-of-time ordered correlator is one effective way to explore the chaotic behavior [11]. By making use of the diagnosis of the chaos, researchers have shown that it increases exponentially with the increase of time [12],i.e., $C(t) \approx e^{\lambda t}$, where the Lyapunov exponent $\lambda$ is used to depict the sensitive dependence on the initial condition of the system [13-15]. Maldacena *et al.* [16, 17] have conjectured the existence of the universal bound from a many-body system on the quantum Lyapunov exponent $\lambda \leq 2\pi T/\hbar$. Here, $T$ is known as the Hawking temperature. It is pointed out that Hawking temperature $T$ can be in direct contact with the black hole's surface gravity $\kappa$ by the relationship $\kappa = 2\pi T$. Naturally, the chaos bound can be rewritten in the form $\lambda \leq \kappa$. In principle, the chaos bound can be tightly associated with the chaotic motion within physical backgrounds of black holes and strongly correlated with many-body systems.

---

* Corresponding author.
E-mail addresses: bingtangphy@jsu.edu.cn

Physically, the single particle case does not need to meet those constraints, which are originated from the many-body system. If one particle feels strong enough scalar or electromagnetic forces, then the particle can be quite close to one black hole but does not fall into it. In this case, the chaotic motion of the particle can be well analyzed. For example, Hashimoto *et al.* have obtained the value of the relevant Lyapunov exponent near the horizon, which is completely consistent with the surface gravity [18]. Their results show that the Lyapunov exponent for one static equilibrium can reach the chaos bound near the black hole horizon, which agrees with the speculation of Ref. [16]. Therefore, even if one single-particle case is considered, the result obtained appears to be in agreement with the chaos bound put forward in Ref. [16] on the quantum thermal system possessing massive degrees of freedom. More relevant theoretical works have displayed the generality of the chaos bound on the particle movement around black holes [19-23]. Particularly, Dalui *et al.* [20] have studied the chaotic behavior of massless and chargeless particle on the black hole near the horizon surface. They have shown that the particle's motion can be chaotic after escaping from the black hole horizon obstacle and the relevant Lyapunov exponent possesses an upper bound determined via the black hole surface gravity. In addition, Čubrović [24] has accomplished an interesting work on the chaotic motion of one closed string in AdS black hole geometries having multiple horizons. They have predicted that the chaos bound can be modified as $\lambda = 2\pi T n$ for the Lyapunov exponent for the string movement, with $n$ being a winding number of the string.

On the other hand, there exists no decoupling case such that near horizon geometry for non-extremal black holes is one particular solution on its own. Hence, the chaos bound is also able to be investigated via analyzing the geodesic movement of the single particle around black holes [25-28]. For example, Zhao *et al.* have considered the near-horizon expansion for the case of one particle at the static equilibrium around the black hole horizon having one applied potential [25]. They have found that the chaos bound can be violated in the situation of the charged particle outside those charged black holes. Furthermore, their results have shown that the particle angular momentum has a remarkable impact on the corresponding Lyapunov exponent by affecting relevant effective potentials and increasing the degree of chaotic motion of one test particle. Recently, a few violations of the chaos bound have been verified in different models and gravitational theories [29-36]. For example, Lei *et al.* have obtained the universal expression for the Lyapunov exponent by considering one charged particle movement around charged black holes, which is an early work to prove that the chaos bound could be violated partially around the Reissner Nordström (RN) and RN-AdS black holes [31]. Straight after, Gao *et al.* [32] have investigated the affecting law of the angular momentum in charged Kiselev black holes around the anisotropic fluid. Their results have shown that the chaos bound can be violated for some particular values of particle angular momentum when the electric charge of the black hole is sufficiently large. These theoretical works play a significant part in the quantum information theory and the black hole physics.

According to Lovelock's theorem [37], if one 4D spacetime, the metricity, the diffeomorphism invariance, and the 2nd order equations of motion are supposed, then general relativity containing one cosmological constant shall be the individual theory for the pure gravity. It is universally accepted that the higher-dimensional Gauss-Bonnet (GB) theory is nontrivial. In 4D spacetime, the GB term in the gravitational action is taken to have no consequences for the field equations of gravity. One fresh theory of gravity in the 4D spacetime is referred to as "4D EGB gravity". This gravity is formulated via rescaling the GB coupling coefficient before the dimensional decrease, which satisfies all previous constraints obtained by steering clear of Lovelock's theorem

[38]. Remarkably, the 4D Einstein-Gauss-Bonnet (EGB) black holes maintain an amount of the graviton degrees of freedom so that it is not affected by the Ostrogradsky instability. Some works have argued that the process of taking the $D \to 4$ limit may be partially compatible [39-42], and the approaches that have resulted from it provide an alternative to the original scheme of Ref.[38]. Since such researches can be done simply for the charged 4D black hole in the EGB theory, more and more attention has been paid to the context of the 4D EGB gravity [43-46]. In current work, we take into account the circular motion of one charged test particle around charged four-dimensional Einstein-Gauss-Bonnet-AdS (4D EGB-AdS) black holes. The influence of both the GB coupling coefficient of the black hole and the angular momentum of the charged particle on the corresponding Lyapunov exponent shall be discussed. Our motivation of the present paper is to investigate the Lyapunov exponent for the charged particle's motion around the 4D EGB-AdS black hole to further comprehend the chaos bound and the black hole chaos phenomenon. We shall try to look for the spatial region in which the chaos bound can be violated. Intriguingly, our results suggest that, in the context of the present black hole, the chaos bound can be violated.

The remaining sections of our article will be organized as follows. In the second section, one solution on 4D EGB-AdS black hole is reviewed. In Section 3, we construct the analytical form for the Lyapunov exponent via considering one eigenvalue problem on the Jacobian matrix in the phase space. In Section 4, we put forward some coefficients to forecast whether the chaos bound of the particle shall be violated in the near horizon expansion. These coefficients can be calculated via taking into account the roles of the GB coupling coefficient and particle angular momentum. In Section 5, we discuss the violation of particle chaos bound in 4D EGB-AdS black holes with different GB coupling coefficients, and look for the spatial region in which particle chaos bound can be violated. Meanwhile, we present violation of particle chaos bound from its circular motion, and obtain the equilibrium orbits for the charged test particle near 4D EGB-AdS black holes. Section 6 will be devoted to summarizing our work.

**2. Charged 4D Einstein-Gauss-Bonnet-AdS black holes**

The action for EGB gravity theory in D dimensions reads

$$S = \frac{1}{16\pi} \int d^D x \sqrt{-g} \left( R - \alpha \mathcal{L}_{GB} \right), \qquad (1)$$

$g$ is the determinant of the metric $g_{\mu\nu}$, $R$ is the Ricci scalar, $\alpha$ is a non-

dimensional coupling coefficient. The $\mathcal{L}_{GB}$ denotes the Lagrangian of the GB gravity, which can be written as

$$\mathcal{L}_{GB} = R_{\mu\nu\lambda\rho}R^{\mu\nu\lambda\rho} - 4R_{\mu\nu}R^{\mu\nu} + R^2, \tag{2}$$

with $R_{\mu\nu}$ the Ricci tensor and $R_{\mu\nu\lambda\rho}$ the Riemann tensor. For the case of $D = 4$, the integral over the Gauss-Bonnet term is a topological invariant, thus it does not contribute to the dynamics [47]. However, as shown in Ref.[38], one can rescale the coupling coefficient as

$$\alpha \to \frac{\alpha}{D-4}. \tag{3}$$

The approach is similar to the dimensional regularization method used in quantum field theory. Consider now the Einstein-Maxwell Gauss-Bonnet theory in D dimensions with a negative cosmological constant given by the action

$$S = \frac{1}{16\pi} \int d^D x \sqrt{-g} \left( R - 2\Lambda + \frac{\alpha}{D-4}\mathcal{L}_{GB} - F_{\mu\nu}F^{\mu\nu} \right), \tag{4}$$

where $\Lambda$ stands for the famous cosmological constant and $F_{\mu\nu} = \partial_\mu A_\nu - \partial_\nu A_\mu$ corresponds to the usual Maxwell tensor. The electromagnetic potential of this black hole has the following form

$$A_\mu = (\frac{Q}{r}, 0, 0, 0). \tag{5}$$

and then taking into account the limit $D \to 4$. Then, Lovelock's theorem is evaded and new spherically symmetric black hole solutions appear.

Here, we focus on the black hole solution depicted with one 4D static spherically symmetric metric, namely,

$$ds^2 = -f(r)dt^2 + \frac{1}{f(r)}dr^2 + r^2(d\theta^2 + \sin^2\theta d\phi^2). \tag{6}$$

If $D \to 4$, then one can write the metric function as [47, 48]

$$f(r) = 1 + \frac{r^2}{2\alpha}\left[1 \pm \sqrt{1 + 4\alpha\left(\frac{2M}{r^3} - \frac{Q^2}{r^4} + \frac{\Lambda}{3}\right)}\right], \tag{7}$$

where $M$ and $Q$ stand for the mass of 4D EGB-AdS black hole and its charge parameters, respectively. The metric function (7) has two branches, which correspond to the selection of $\pm$. Considering that "+" branch cannot produce one physically meaningful solution [49, 50], hence we shall restrict the present work to "-" branch of

Eq.(7). The outer event horizon of 4D EGB-AdS black holes $r_+$ corresponds to a larger root of the high-power equation, which has the following form

$$1 - \frac{2M}{r} + \frac{Q^2 + \alpha}{r^2} - \frac{\Lambda}{3} r^2 = 0. \tag{8}$$

At the outer horizon of the present charged black hole, its surface gravity can be written as

$$\kappa = -\frac{1}{\sqrt{g_{rr}}} \frac{d\sqrt{g_{tt}}}{dr}\bigg|_{r=r_+} = \frac{1}{2} f(r)'\bigg|_{r=r_+}. \tag{9}$$

Here, the prime " ′ " stands for the first order derivative with respect to $r$. If we neglect the cosmological constant, then Eq.(8) has the following simple solution

$$r_\pm = M \pm \sqrt{M^2 - Q^2 - \alpha}. \tag{10}$$

When there exists one nonzero cosmological constant, the expression of $r_+$ is very extremely tedious and does not have a certain form, so it is not displayed here. The physical natures of the "-" branch differ relying on whether the black hole mass $M$ is smaller or larger than one threshold mass, which is

$$M_* = \sqrt{Q^2 + \alpha}. \tag{11}$$

In Fig. 1, the radial dependence of metric coefficient $f(r)$ for the "-" branch is displayed by considering three different cases: (1) $M < M_*$, (2) $M > M_*$ and (3) $M = M_*$ [47]. If $M < M_*$ there do not exist any horizons, i.e., the black hole solution is not allowed to appear. Once $M > M_*$ then there exist two black hole horizons, as presented in Eq.(10). When $M = M_*$ one can find a degenerate horizon, which corresponds to one extremal black hole.

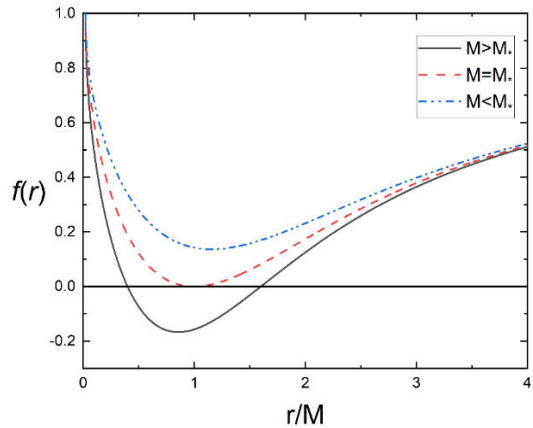

**Fig. 1.** The metric coefficient $f(r)$ as one function of the radial position for the "-" branch. Here, the cosmological constant has been set to be zero.

## 3. The charged particle motion around the black hole

The Lagrangian for one charged test particle around one black hole reads

$$\mathcal{L} = \frac{1}{2}g_{\mu\nu}\dot{X}_\mu \dot{X}_\nu - \frac{q}{m}A_\mu \dot{X}_\mu. \tag{12}$$

Here, dot signifies the first-order derivative with respect to $\tau$, $q$ corresponds to the charge of one test particle, $m$ is its mass, $g_{\mu\nu}$ is called the metric tensor, and $A_\mu$ denotes electromagnetic vector potential. For the sake of analysis, we fix $m=1$ throughout the whole article.

When one probe particle with charge $q$ moves on the equatorial plane ($\theta = \frac{\pi}{2}$, $\dot{\theta}=0$) of the charged black hole, relevant Lagrangian can be organized into

$$\mathcal{L} = \frac{1}{2}(-f\dot{t}^2 + \frac{1}{f}\dot{r}^2 + r^2\dot{\phi}^2) - \frac{qQ}{r}\dot{t}. \tag{13}$$

Physically, the generalized momenta can be defined via $P_\mu = \partial \mathcal{L}/\partial \dot{X}_\mu$. Doing some simple calculations, we find that these generalized momenta read

$$P_t = -f\dot{t} - \frac{qQ}{r} = -E, \tag{14}$$

$$P_r = \frac{\dot{r}}{f}, \tag{15}$$

$$P_\phi = r^2\dot{\phi} = L. \tag{16}$$

By considering $H = P_\mu \dot{X}^\mu - \mathcal{L}$, one can construct the Hamiltonian of the charged particle, which can be expressed as

$$H = \frac{-(P_t + \frac{qQ}{r})^2 + P_r^2 f^2 + P_\phi^2 r^{-2} f}{2f}. \tag{17}$$

Taking advantage of the canonical equation, one shall get the equation of motions, namely,

$$\dot{t} = -\frac{P_t + \frac{qQ}{r}}{f}, \dot{P_t} = 0, \dot{r} = P_r f,$$

$$\dot{P_r} = -\frac{1}{2}[2(P_t + \frac{qQ}{r})\frac{qQ}{fr^2} + (P_t + \frac{qQ}{r})^2 \frac{f'}{f^2} + P_r^2 f' - 2P_\phi^2 r^{-3}], \qquad (18)$$

$$\dot{\phi} = \frac{P_\phi}{r^2}, \dot{P_\phi} = 0$$

Reorganizing equations of motion in the coordinate time $t$, we have

$$\frac{dr}{dt} = -\frac{P_r f^2}{P_t + \frac{qQ}{r}} = F_1, \qquad (19)$$

$$\frac{dP_r}{dt} = \frac{qQ}{r^2} + \frac{1}{2}[\frac{(P_t + \frac{qQ}{r})f'}{f} + \frac{P_r^2 f'f}{P_t + \frac{qQ}{r}} - \frac{2P_\phi^2 r^{-3} f}{P_t + \frac{qQ}{r}}] = F_2. \qquad (20)$$

The four-velocity of a massive particle must obey the normalization condition $g_{\mu\nu}\dot{X}_\mu \dot{X}_\nu = -1$. Thus, one can get the restricted condition

$$P_t = -\frac{qQ}{r} - \sqrt{f(1 + P_r^2 f + r^{-2} P_\phi^2)}. \qquad (21)$$

Theoretically, the Lyapunov exponent can be calculated via working out the eigenvalue problem on the related Jacobian matrix in one phase space $(r, P_r)$. Here, the Jacobian matrix can be defined via $K_{ij}$, which have the following four elements

$$K_{11} = \frac{\partial F_1}{\partial r}, K_{12} = \frac{\partial F_1}{\partial P_r}, K_{21} = \frac{\partial F_2}{\partial r}, K_{22} = \frac{\partial F_2}{\partial P_r}. \qquad (22)$$

In the present work, the circling motion of the charged test particle is restricted on the equatorial plane in charged 4D EGB-AdS black holes. When the charged test particle is in equilibrium, we note $P_r = \frac{dP_r}{dt} = 0$. In this case, the Jacobian matrix $K_{ij}$ can be simplified as

$$K_{11} = K_{22} = 0,$$

$$K_{12} = f\sqrt{\frac{f}{1 + r^{-2}P_\phi^2}}, \qquad (23)$$

$$K_{21} = -2\frac{qQ}{r^3} + \frac{[-6P_\phi^2 r^{-4} f + 4P_\phi^2 r^{-3} f' - (1 + r^{-2}P_\phi^2)f'']}{2\sqrt{f(1 + r^{-2}P_\phi^2)}} + \frac{[f'(1 + r^{-2}P_\phi^2) - 2fr^{-3}P_\phi^2]^2}{4f(1 + r^{-2}P_\phi^2)\sqrt{f(1 + r^{-2}P_\phi^2)}}.$$

Since the Lyapunov exponent for a test particle in the circular orbit can be derived

via working out the eigenvalue problem on the Jacobian matrix, we have

$$\lambda^2 = K_{12}K_{21}$$
$$= \frac{\left[f'(1+r^{-2}P_\phi^2)-2fr^{-3}P_\phi^2\right]^2}{4(1+r^{-2}P_\phi^2)^2} - \frac{2f^2}{\sqrt{f(1+r^{-2}P_\phi^2)}}\frac{qQ}{r^3} - \frac{\left[6P_\phi^2 r^{-4}f - 4P_\phi^2 r^{-3}f' + (1+r^{-2}P_\phi^2)f''\right]f}{2(1+r^{-2}P_\phi^2)}. \quad (24)$$

By combining the restricted condition $P_r = \frac{dP_r}{dt} = 0$ with Eq.(20), one can obtain

$$\frac{qQ}{r^2} + \frac{2P_\phi^2 r^{-3}f - (1+r^{-2}P_\phi^2)f'}{2\sqrt{f(1+r^{-2}P_\phi^2)}} = 0. \quad (25)$$

There exists one static equilibrium circumstance on the charge $q$ of the massive particle at the balance position, which is

$$q = -\frac{2fP_\phi^2 r^{-1} - (1+r^{-2}P_\phi^2)f'r^2}{2Q\sqrt{f(1+r^{-2}P_\phi^2)}}\bigg|_{r=r_0}. \quad (26)$$

Therefore, substituting (26) into (24), one can find that the Lyapunov exponent meets the following relation

$$\lambda^2 = \frac{1}{4}\left[f'^2 - \frac{4P_\phi^2 f^2}{(r^2+P_\phi^2)^2} - f\left(\frac{4rf'}{r^2+P_\phi^2} + 2f''\right)\right], \quad (27)$$

Considering that the existence of the particle angular momentum is very important to the Lyapunov exponent, hence it will not be neglected in the present work. It is not hard to verify that Eq.(27) can reduce to $\lambda^2 = \frac{1}{4}f'^2 = \kappa^2$ when $f(r_+) = 0$. This means that the chaos bound can be saturated at the black hole's outer horizon.

## 4. Near-horizon geometry behavior analysis

Theoretically, near horizon geometries of the 4D black hole shall provide one more distinct physical explanation. In fact, this geometry can be specified via the metric function. Its Taylor expansion at the outer horizon $r = r_+$ has a series form [25, 31]

$$f(r) = f_1(r-r_+) + f_2(r-r_+)^2 + f_3(r-r_+)^3 + \cdots. \quad (28)$$

For the sake of further analyzing whether the chaos bound can be violated via circling motion of one charged particle, we insert Eq.(28) into Eq.(9) and Eq.(27), which yields

$$\kappa = \frac{1}{2}\Big[f_1 + 2f_2(r-r_+) + 3f_3(r-r_+)^2\Big]\Big|_{r=r_+} = \frac{1}{2}f_1, \tag{29}$$

$$\begin{aligned}\lambda^2 =& \frac{1}{4}\Big[f_1 + 2f_2(r-r_+) + 3f_3(r-r_+)^2\Big]^2 \\ & -\frac{(r-r_+)^2 P_\phi^2}{(r^2+P_\phi^2)^2}\Big[f_1 + f_2(r-r_+) + f_3(r-r_+)^2\Big]^2 \\ & -\Big[f_2 + 3f_3(r-r_+)\Big]\Big[f_1(r-r_+) + f_2(r-r_+)^2 + f_3(r-r_+)^3\Big] \\ & -\frac{r}{r^2+P_\phi^2}\Big[f_1 + 2f_2(r-r_+) + 3f_3(r-r_+)^2\Big]\Big[f_1(r-r_+) + f_2(r-r_+)^2 + f_3(r-r_+)^3\Big]\end{aligned} \tag{30}$$

Then, the Lyapunov exponent $\lambda$ in the vicinity of $r_+$ can be expanded into the following form

$$\lambda^2 = \kappa^2 + \gamma_1(r-r_+) + \gamma_2(r-r_+)^2 + \sum_{n=3}^{\infty}\gamma_n(r-r_+)^n, \tag{31}$$

Here, $\gamma_1$ and $\gamma_2$ are two parameters, which are respectively defined as

$$\gamma_1 = -\frac{r_+ f_1^2}{r_+^2 + P_\phi^2}, \tag{32}$$

$$\gamma_2 = \frac{f_1^2 - 3r_+ f_1 f_2}{r_+^2 + P_\phi^2} - \frac{3P_\phi^2 f_1^2}{(r_+^2 + P_\phi^2)^2} - \frac{3f_1 f_3}{2}. \tag{33}$$

Their signs jointly determine whether $\kappa$ is viewed as one lower bound or one upper bound. The connection between Lyapunov exponent $\lambda$ and black hole surface gravity $\kappa$ has the following form

$$\begin{cases}\gamma_1(r-r_+) + \gamma_2(r-r_+)^2 + \sum_{n=3}^{\infty}\gamma_n(r-r_+)^n < 0, & \lambda < \kappa \\ \gamma_1(r-r_+) + \gamma_2(r-r_+)^2 + \sum_{n=3}^{\infty}\gamma_n(r-r_+)^n = 0, & \lambda = \kappa \\ \gamma_1(r-r_+) + \gamma_2(r-r_+)^2 + \sum_{n=3}^{\infty}\gamma_n(r-r_+)^n > 0, & \lambda > \kappa\end{cases} \tag{34}$$

From Eq.(31), we can see that the coefficients $\gamma_1$ and $\gamma_2$ can control whether the circular motion of the test particle within near horizon geometry shall violate the chaos bound. To better conclude the Lyapunov exponential behavior of circular motion, we go into detail about $\gamma_1$ and $\gamma_2$. Here, we choose the values of the coefficients of charged 4D EGB-AdS black holes ($M=1$, $Q=0.2$, $\Lambda=-0.02$) so as to reduce subsequent calculations. In Fig. 2, we show the relationship between $\gamma_i (i=1,2)$ and the GB coupling coefficient of 4D EGB-AdS black holes, while Fig. 3 discusses their relationship with the particle angular momentum. In both Fig. 2(a) and Fig. 3(a), we can clearly see that the $\gamma_1$ is negative and approaches 0 with the increase of GB coupling coefficient $\alpha$ or particle angular momentum. However, $\gamma_2$ changes in two ways with the increase of $\alpha$: (i) in the case of $L = 3 \sim 30$, it increases first and remains greater than 0 after passing through the axis at point $A$; (ii) for the case of small angular momentum $L = 0, 1$, it decreases first and remains less than zero after passing through the axis at point $B$. We note that no violation exists when $\alpha < \alpha_A$ for the case of $L = 3 \sim 30$, and the small particle angular momentum (about between 0 and 3) of the circular motion shall not go over the chaos bound when $\alpha > \alpha_B$. Fig. 3(b) shows that $\gamma_2$ is positive for the case $\alpha = 0.3 \sim 0.8$ but negative for the case $\alpha = 0.1, 0.2$. Particularly, the change of $\gamma_2$ is rapidly complex in the interval $0 < L < 3$, and then it tends to one specific constant value. There exists a reasonable concordance between the two sets of results about small angular momentum.

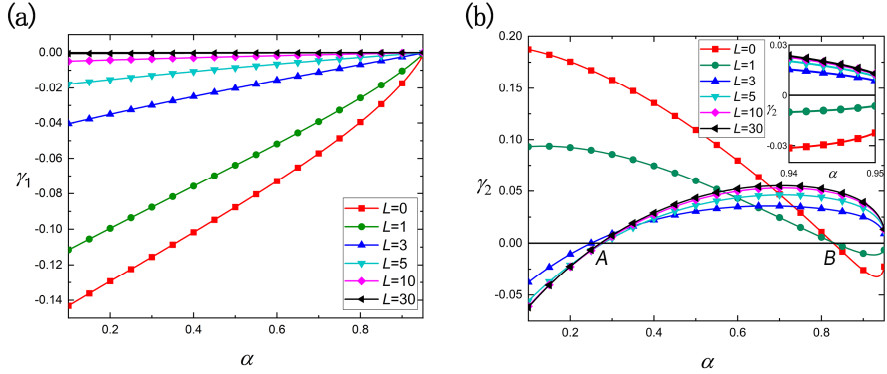

**Fig. 2.** The parameters $\gamma_1$, $\gamma_2$ as one function of the GB coupling coefficient $\alpha$ for various particle angular momentum.

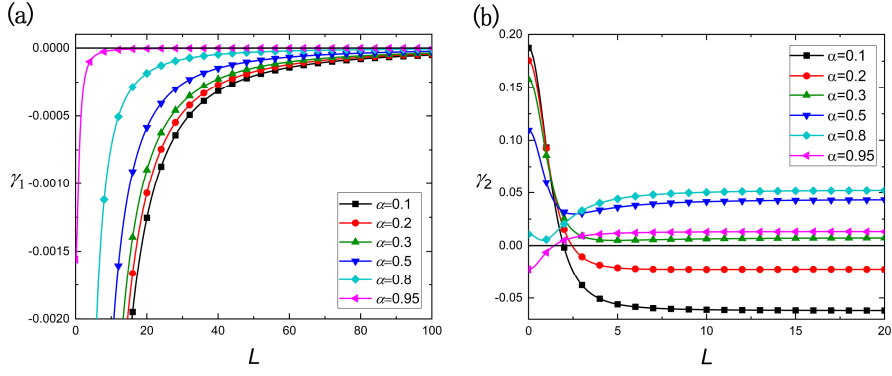

**Fig. 3.** The parameters $\gamma_1$, $\gamma_2$ as one function of particle angular momentum $L$ for various GB coupling coefficients of black holes.

To check whether the circling motion of the charged test particle can transcend the chaos bound, let us analyze the influence of the expansion part $\gamma_1(r-r_+) + \gamma_2(r-r_+)^2$. We define a ratio as $\Delta = \dfrac{\gamma_2(r-r_+)^2}{\gamma_1(r-r_+)}$ to control whether the total contribution of the expansion term is regulated via $\gamma_1$ or $\gamma_2$. When $|\Delta| > 1$, the sign of expansion term is defined by $\gamma_2$, or else by $\gamma_1$. We can introduce $\delta = r_0 - r_+$ to simplify $\Delta = \dfrac{\gamma_2 \delta}{\gamma_1}$, where $\delta$ is a positive small quantity obeying the near-horizon expansion. As the GB coupling coefficient or angular momentum increases, the value of $\gamma_1$ tends to zero. Thus, we can obtain $\delta \gg \gamma_1$ when the particle angular momentum is sufficiently large, which will lead to $|\Delta| > 1$. At this point, the key is whether $\gamma_2$ is positive or negative. Once $\gamma_2$ is negative, the Lyapunov exponent for test particle circular motion agrees with the chaos bound. When the GB coupling coefficient is large enough, the $\gamma_2$ is positive for $L = 3 \sim 30$, which indicates the chaos bound can be violated for the particle's circular motion. Hence, one can infer that the particle circular motion possessing a sufficiently large angular momentum ($L = 3 \sim 30$) shall violate the chaos bound in the near horizon region of 4D EGB-AdS black hole ($\alpha > 0.3$).

## 5. The Lyapunov exponent analysis

It should be noted that the above near horizon expansion is not a distinctive situation to estimate the contravention of the chaos bound. Once an analytical form of the Lyapunov exponent $\lambda$ is gotten, one can compare the values of $\lambda$ and $\kappa$ in all region outside black hole horizon. The emphasis is on boundary violation, which can occur for a sufficiently large GB coupling coefficient. The particle angular momentum is very important to the investigation of those factors on boundary. While the charge of test particle is directly fixed and the GB coupling coefficient is sufficiently large, the chaos bound shall be violated for some particular values of the particle angular momentum.

5.1. Beyond the near horizon region

Through deriving the parameters ($\gamma_1$, $\gamma_2$), we forecast that circular motion of test particle shall violate the chaos bound. For the sake of supporting the above theoretical results, our analysis shall be extended to a circumstance beyond the near-horizon region. Sine $r_0$ is a parameter of the Lyapunov exponent, one can get the full expression of the $\lambda^2$ at every position of the radial coordinate. With Eq.(27), the correlation curve between the Lyapunov exponent and radial coordinate $r_0$ can be plotted, as shown in Fig. 4. Obviously, they are in agreement with the near horizon results of $\lambda^2$. In Fig. 4, one can clearly see that the square of the Lyapunov exponent of the charged test particle in the static equilibrium is always under the line of the surface gravity $\kappa^2$. This reveals that the existence of angular momentum causes a larger Lyapunov exponent for the particle's circular motion, which leads to the contravention of the chaos bound. With Eq.(26), one can evaluate the charge value of the particle, which corresponds to the violation of the chaos bound. Relevant results are shown in Table 1.

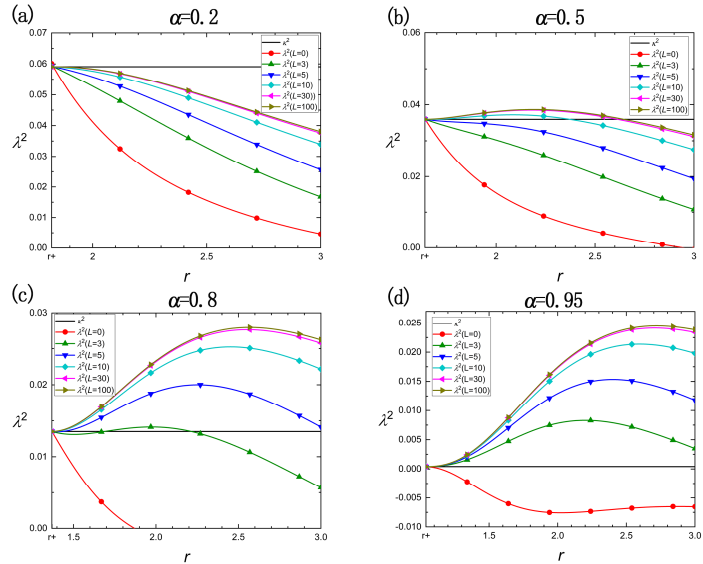

**Fig. 4.** The Lyapunov exponents $\lambda^2$ in Eq.(27) as a function of $r$ for various GB coupling coefficients of the charged black hole.

The GB coupling coefficient of the present black hole can affect the chaos bound. As displayed in Fig. 4(a), when values of the electric charge, the mass and the cosmological constant of black holes are fixed, and let $\alpha = 0.2$, no matter how the value of particle angular momentum varies, the chaos bound cannot be violated. According to Figs. 4(b)-(d), we note that the chaos bound is more easily violated when the other black holes properties parameters are fixed and the closer $\alpha$ is to 0.96. When the GB coupling coefficient is sufficiently large, the circular motion of the test particle violates the chaos bound at some particular values of the radial coordinate. Moreover, the Lyapunov exponent always meets $\lambda^2 = \kappa^2$ for the horizon. Thus, the chaos bound $\lambda^2 \leq \kappa^2$ can be saturated at the event horizon. It is shown that, as the GB coupling coefficient increases, radial coordinate and charge intervals (corresponding to the contravention of the chaos bound) increase, but the minimum value of test particle charge decreases. We can let the orbit of the test particle infinitely approaches the outer horizon when the charge/mass ratio is sufficiently large. Theoretically, the event horizon can be viewed as a nest of chaos.

**Table 1**

The radial position of the equilibrium orbit and the corresponding charge of the test particle with the angular momentum $L = 50$ for various values of the GB coupling coefficient.

| $\lambda^2 > \kappa^2$ | $r$ | $q$ |
|---|---|---|
| $\alpha = 0.3$ | [1.798, 1.857] | [466, 900] |
| $\alpha = 0.5$ | [1.644, 2.635] | [13, 2708] |
| $\alpha = 0.8$ | [1.371, 4.505] | [-104, 3476] |

5.2. Chaos violation from probe particle's circular motion

Let us consider one test particle's circular motion around the charged 4D EGB-AdS black hole to analyze the chaos bound. Due to the complexity of the analytical forms of the equilibrium orbits, the positions $r_0$ of some certain need to be numerically obtained. The corresponding dates are shown in Table 2. Here, we set $m = 1$, $q = 15$. When the charge, cosmological constant and the particle angular momentum are directly fixed, the values of positions of the orbit and the black hole horizon decrease as the value of the GB coupling coefficient increases. On the other part, we note that the position of the orbit gradually shifts away from the horizon as increases particle angular momentum, which agrees with previous results[32]. Therefore, we can control the GB coupling coefficient and angular momentum to cause that charged particle's orbit to be arbitrarily closed to the black hole horizon.

**Table 2**

The radial position of the equilibrium orbit of probe particle with various angular momentums, $L=1$, $L=30$ and $L=100$, for various values of the GB coupling coefficient.

| $\alpha$ | 0.1 | 0.2 | 0.5 | 0.8 | 0.95 |
|---|---|---|---|---|---|
| $r_0(L=1)$ | 2.12781 | 2.04385 | 1.76341 | 1.40880 | 1.04795 |
| $r_0(L=30)$ | 2.76808 | 2.71832 | 2.54599 | 2.31505 | 2.15149 |
| $r_0(L=100)$ | 2.87601 | 2.82969 | 2.67124 | 2.46632 | 2.33045 |

The natures of the black hole can obviously affect the chaos bound. Fig. 5 shows the results of the variation of $\lambda^2 - \kappa^2$ with $L$ corresponding to different values of the GB coupling coefficients. If values of electric charge and cosmological constant are directly fixed, then the value of $\lambda^2 - \kappa^2$ converges to a certain constant value with increasing angular momentum, and increases as the GB coupling coefficient increases. It is more likely to lead to the violation of the bound while the value of the GB coupling coefficient is relatively larger. We find that the chaos bound is not violated in such regions when $\alpha = 0.3$. This is because the charge of the particle needs to satisfy $q > 466$ as $\alpha = 0.3$ case in Table 1, which is far from the condition of $q = 15$. These results accord closely with our predictions.

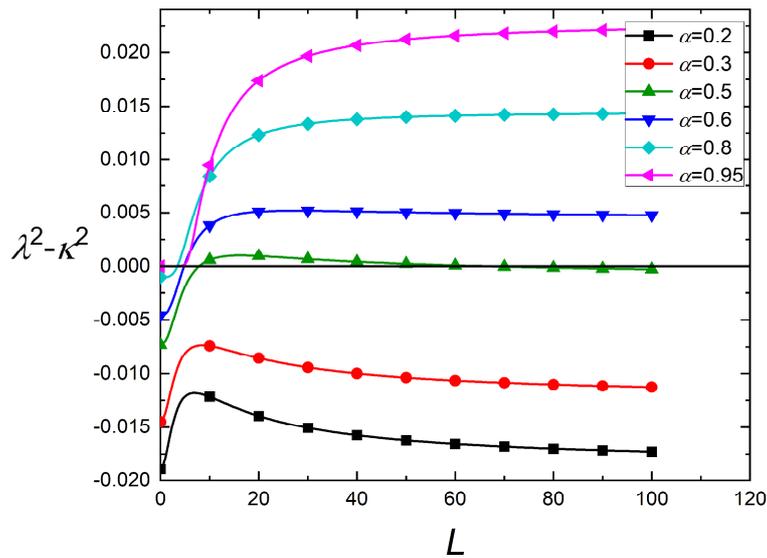

**Fig. 5.** The effect of the particle angular momentum on the chaos bound for various values of the GB coupling coefficient.

**6. Conclusion**

In summary, we have investigated the Lyapunov exponent for one charged test particle in near horizon geometry and beyond the near-horizon region of 4D charged Einstein-Gauss-Bonnet-AdS black holes to identify the validation of the chaos bound. The Lyapunov exponent for the particle's circular motion around 4D EGB-AdS black holes has been calculated via making use of the Jacobian matrix. Through a general, we have found that the circular motion of one probe particle at black hole horizon can saturate the chaos bound. We have defined some coefficients to connect the relationship between the Lyapunov exponent $\lambda$ and the black hole surface gravity $\kappa$. By expanding the Lyapunov exponent near the horizon and calculating the analytical form of the Lyapunov exponent $\lambda$, we have showed that it has some specific values to determine whether a violation of chaos bound. When the charge, cosmological constant and particle angular momentum are directly fixed, values of positions of the orbit and black hole horizon decrease as the GB coupling coefficient increase, and the chaos is indeed strengthened. Our result shows that the presence of particle angular momentum generates the violation of the chaos bound while the GB coupling coefficient is sufficiently large. We have especially displayed that as the GB coupling coefficient increases, the value of the charge of the test particle violating the chaos bound is even smaller.

**Acknowledgments**

The present theoretical work was supported by the National Natural Science Foundation of China under Grant Nos. 11875126 and 11964011.